\documentclass[ALICE,manyauthors]{cernphprep}
\usepackage[comma,square,numbers,sort&compress]{natbib}

\usepackage{lineno}
\usepackage{xspace}
\usepackage{hyperref}
\usepackage{color}
\usepackage[T1]{fontenc}
\usepackage{orcidlink}

\newcommand{\hyp}{${}^3_{\Lambda}\mathrm{H}$}

\newcommand{\hyper}{${}^4_{\Lambda}\mathrm{H}$}
\newcommand{\ahyper}{${}^4_{\bar{\Lambda}}\overline{\mathrm{H}}$}
\newcommand{\hyperE}{${}^4_{\Lambda}\mathrm{He}$}
\newcommand{\ahyperE}{${}^4_{\bar{\Lambda}}\overline{\mathrm{He}}$}

\begin{document}
%

\newcommand{\pp}           {pp\xspace}
\newcommand{\ppbar}        {\mbox{$\mathrm {p\overline{p}}$}\xspace}
\newcommand{\XeXe}         {\mbox{Xe--Xe}\xspace}
\newcommand{\PbPb}         {\mbox{Pb--Pb}\xspace}
\newcommand{\pA}           {\mbox{pA}\xspace}
\newcommand{\pPb}          {\mbox{p--Pb}\xspace}
\newcommand{\AuAu}         {\mbox{Au--Au}\xspace}
\newcommand{\dAu}          {\mbox{d--Au}\xspace}

\newcommand{\s}            {\ensuremath{\sqrt{s}}\xspace}
\newcommand{\snn}          {\ensuremath{\sqrt{s_{\mathrm{NN}}}}\xspace}
\newcommand{\pt}           {\ensuremath{p_{\rm T}}\xspace}
\newcommand{\meanpt}       {$\langle p_{\mathrm{T}}\rangle$\xspace}
\newcommand{\ycms}         {\ensuremath{y_{\rm CMS}}\xspace}
\newcommand{\ylab}         {\ensuremath{y_{\rm lab}}\xspace}
\newcommand{\etarange}[1]  {\mbox{$\left | \eta \right |~<~#1$}}
\newcommand{\yrange}[1]    {\mbox{$\left | y \right |~<~#1$}}
\newcommand{\dndy}         {\ensuremath{\mathrm{d}N_\mathrm{ch}/\mathrm{d}y}\xspace}
\newcommand{\dndeta}       {\ensuremath{\mathrm{d}N_\mathrm{ch}/\mathrm{d}\eta}\xspace}
\newcommand{\avdndeta}     {\ensuremath{\langle\dndeta\rangle}\xspace}
\newcommand{\dNdy}         {\ensuremath{\mathrm{d}N_\mathrm{ch}/\mathrm{d}y}\xspace}
\newcommand{\Npart}        {\ensuremath{N_\mathrm{part}}\xspace}
\newcommand{\Ncoll}        {\ensuremath{N_\mathrm{coll}}\xspace}
\newcommand{\dEdx}         {\ensuremath{\textrm{d}E/\textrm{d}x}\xspace}
\newcommand{\RpPb}         {\ensuremath{R_{\rm pPb}}\xspace}

\newcommand{\nineH}        {$\sqrt{s}~=~0.9$~Te\kern-.1emV\xspace}
\newcommand{\seven}        {$\sqrt{s}~=~7$~Te\kern-.1emV\xspace}
\newcommand{\twoH}         {$\sqrt{s}~=~0.2$~Te\kern-.1emV\xspace}
\newcommand{\twosevensix}  {$\sqrt{s}~=~2.76$~Te\kern-.1emV\xspace}
\newcommand{\five}         {$\sqrt{s}~=~5.02$~Te\kern-.1emV\xspace}
\newcommand{\twosevensixnn}{$\sqrt{s_{\mathrm{NN}}}~=~2.76$~Te\kern-.1emV\xspace}
\newcommand{\fivenn}       {$\sqrt{s_{\mathrm{NN}}}~=~5.02$~Te\kern-.1emV\xspace}
\newcommand{\LT}           {L{\'e}vy-Tsallis\xspace}
\newcommand{\GeVc}         {Ge\kern-.1emV/$c$\xspace}
\newcommand{\MeVc}         {Me\kern-.1emV/$c$\xspace}
\newcommand{\TeV}          {Te\kern-.1emV\xspace}
\newcommand{\GeV}          {Ge\kern-.1emV\xspace}
\newcommand{\MeV}          {Me\kern-.1emV\xspace}
\newcommand{\GeVmass}      {Ge\kern-.2emV/$c^2$\xspace}
\newcommand{\MeVmass}      {Me\kern-.2emV/$c^2$\xspace}
\newcommand{\lumi}         {\ensuremath{\mathcal{L}}\xspace}

\newcommand{\ITS}          {\rm{ITS}\xspace}
\newcommand{\TOF}          {\rm{TOF}\xspace}
\newcommand{\ZDC}          {\rm{ZDC}\xspace}
\newcommand{\ZDCs}         {\rm{ZDCs}\xspace}
\newcommand{\ZNA}          {\rm{ZNA}\xspace}
\newcommand{\ZNC}          {\rm{ZNC}\xspace}
\newcommand{\SPD}          {\rm{SPD}\xspace}
\newcommand{\SDD}          {\rm{SDD}\xspace}
\newcommand{\SSD}          {\rm{SSD}\xspace}
\newcommand{\TPC}          {\rm{TPC}\xspace}
\newcommand{\TRD}          {\rm{TRD}\xspace}
\newcommand{\VZERO}        {\rm{V0}\xspace}
\newcommand{\VZEROA}       {\rm{V0A}\xspace}
\newcommand{\VZEROC}       {\rm{V0C}\xspace}
\newcommand{\Vdecay} 	   {\ensuremath{V^{0}}\xspace}

\newcommand{\ee}           {\ensuremath{e^{+}e^{-}}} 
\newcommand{\pip}          {\ensuremath{\pi^{+}}\xspace}
\newcommand{\pim}          {\ensuremath{\pi^{-}}\xspace}
\newcommand{\kap}          {\ensuremath{\rm{K}^{+}}\xspace}
\newcommand{\kam}          {\ensuremath{\rm{K}^{-}}\xspace}
\newcommand{\pbar}         {\ensuremath{\rm\overline{p}}\xspace}
\newcommand{\kzero}        {\ensuremath{{\rm K}^{0}_{\rm{S}}}\xspace}
\newcommand{\lmb}          {\ensuremath{\Lambda}\xspace}
\newcommand{\almb}         {\ensuremath{\overline{\Lambda}}\xspace}
\newcommand{\Om}           {\ensuremath{\Omega^-}\xspace}
\newcommand{\Mo}           {\ensuremath{\overline{\Omega}^+}\xspace}
\newcommand{\X}            {\ensuremath{\Xi^-}\xspace}
\newcommand{\Ix}           {\ensuremath{\overline{\Xi}^+}\xspace}
\newcommand{\Xis}          {\ensuremath{\Xi^{\pm}}\xspace}
\newcommand{\Oms}          {\ensuremath{\Omega^{\pm}}\xspace}
\newcommand{\degree}       {\ensuremath{^{\rm o}}\xspace}

\begin{titlepage}
\PHyear{2024}       
\PHnumber{265}      
\PHdate{08 October}  

\title{First measurement of A = 4 (anti)hypernuclei at the LHC}
\ShortTitle{First measurement of A = 4 (anti)hypernuclei}   

\Collaboration{ALICE Collaboration\thanks{See Appendix~\ref{app:collab} for the list of collaboration members}}
\ShortAuthor{ALICE Collaboration} 

\begin{abstract}
 In this Letter, the first evidence of the \ahyperE~antihypernucleus is presented, along with the first measurement at the LHC of the production of (anti)hypernuclei with mass number $A=4$, specifically (anti)\hyper~ and (anti)\hyperE. In addition, the antiparticle-to-particle ratios for both hypernuclei (\ahyper{} / \hyper ~and \ahyperE{} / \hyperE) are shown, which are sensitive to the baryochemical potential of the strongly-interacting matter created in heavy-ion collisions. The results are obtained from a data sample of central Pb--Pb collisions, collected during the 2018 LHC data-taking at a center-of-mass energy per nucleon pair of $\sqrt{s_{\mathrm{NN}}} = $ 5.02 TeV. The yields measured for the average of the charge-conjugated states  are found to be $[0.78 \; \pm \; 0.19  \; \mathrm{(stat.)} \; \pm \; 0.17 \; \mathrm{(syst.)}]  \times 10^{-6}$ for the (anti)\hyper ~and \mbox{$[1.08 \; \pm \; 0.34  \; \mathrm{(stat.)} \; \pm \; 0.20  \; \mathrm{(syst.)}]  \times 10^{-6}$} for the (anti)\hyperE, and the measured antiparticle-to-particle ratios are in agreement with unity. The presence of (anti)\hyper~and (anti)\hyperE~excited states is expected to strongly enhance the production yield of these hypernuclei. The yield values exhibit a combined deviation of 3.3$\sigma$ from the theoretical ground-state-only expectation, while the inclusion of the excited states in the calculations leads to an agreement within 0.6$\sigma$ with the present measurements. Additionally, the measured (anti)\hyper ~and (anti)\hyperE ~masses are compatible with the world-average values within the uncertainties. 

\end{abstract}
\end{titlepage}

\setcounter{page}{2} 


Hypernuclei, objects composed of nucleons and hyperons, provide unique insights into the forces that bind strange hadrons with ordinary matter~\cite{RevMod88:Strangeness, massshift, STAR:2021orx}. These objects decay weakly after a few hundred picoseconds into two or more decay products ~\cite{RevMod88:Strangeness}. An understanding of hyperon and nucleon interactions inside of hypernuclei can be used to constrain the models describing the equation of state of neutron stars~\cite{Djapo:2008au,Rezzolla:2018jee,Logoteta:2019utx,Tolos:2020aln}, though to-date these interactions are theoretically poorly understood and have rarely been measured. One way to produce hypernuclei is through ultrarelativistic heavy-ion collisions, carried out at RHIC at Brookhaven National Laboratory or at the Large Hadron Collider (LHC) at CERN~\cite{Braun-Munzinger:2018hat, STAR:2023fbc, STAR:2021orx, ALICE:2015oer, Davis:2005mb}, in which the production yields per nucleon--nucleon collision are enhanced with respect to those in pp collisions. The lightest and most abundant hypernucleus produced in heavy-ion collisions is the hypertriton (\hyp), which can be described as a bound state of a deuteron and a $\Lambda$ hyperon, having a $\Lambda$-separation energy of only about 100 keV~\cite{Braun-Munzinger:2018hat,Hildenbrand:2019sgp,Hildenbrand:2023rkc, massshift}. The discovery of its antihypernucleus by the STAR Collaboration in ultrarelativistic heavy-ion collisions at RHIC~\cite{Abelev:2010} was an important experimental achievement. Recently, the STAR Collaboration also published the first observation of the antihyperhydrogen-4 (\ahyper)~\cite{STAR:2023fbc}, the bound state of an antiproton, two antineutrons, and a $\overline{\Lambda}$. 

This Letter reports the first evidence of the \mbox{antihyperhelium-4~(\ahyperE)}, and in addition, the first measurement of the production yields of (anti)hypernuclei with mass number $A=4$, namely (anti)\hyper, and (anti)\hyperE,~at the LHC. The (anti)\hyper ~is reconstructed from its charged mesonic two-body decay into an (anti)${}^{4}$He and a charged pion, while the \mbox{(anti)\hyperE}~is reconstructed from its charged mesonic three-body decay into an (anti)${}^{3}$He, an (anti)proton, and a charged pion~\cite{RevMod88:Strangeness}. Both (anti)hypernuclei have $\Lambda$-separation energies of about two MeV~\cite{J-PARCE13:2015uwb}, which results in compact hypernuclear structures. 

The measurement of hypernuclei in heavy-ion collisions can also provide information on the bulk properties of the medium created. In such a heavy-ion collision, a strongly-interacting quark--gluon plasma (QGP) is formed~\cite{Braun-Munzinger:2007edi}, composed of deconfined quarks and gluons. The QGP expands and cools down until it reaches the hadronization temperature, at which point it transitions to a system composed of hadrons and antihadrons. The production of hadrons and (anti)nuclei in heavy-ion collisions can be well described by the Statistical Hadronization Model (SHM) over nine orders of magnitude and with an accuracy of about 10\%~\cite{Braun-Munzinger:2007edi, Andronic:2010qu, Andronic:2017pug, Vovchenko:2018fiy}. The SHM utilizes a thermal description of the production of (anti)particles, which depends on only three free parameters, the chemical freeze-out temperature (\textit{T}$_{\mathrm{ch}}$), the volume of the system, and the baryochemical potential. This means that the yield of a given hadron or nucleus species depends only on its mass and spin degeneracy, while (anti)nuclei, having a composite structure, are treated like hadrons.
\textit{T}$_{\mathrm{ch}}$ is the temperature of the particle-emitting source at the point where inelastic interactions among the hadrons cease and the relative abundances of the different hadron species are frozen. At LHC energies, \textit{T}$_{\mathrm{ch}}$ is found to be about 155 MeV~\cite{Andronic:2017pug, Vovchenko:2018fiy}, very close to the (pseudo)critical temperature for the transition from the QGP to a hadron gas obtained from quantum chromodynamics calculations on the lattice~\cite{HotQCD:2018pds, Borsanyi:2020fev}. The baryochemical potential is proportional to the net baryon density of the system and can be determined by measuring antiparticle-to-particle ratios. At LHC energies, the baryochemical potential is very close to zero at midrapidity, meaning that baryons and antibaryons are produced at equal amounts and their ratio is close to unity~\cite{ALICE:2023ulv}. (Anti)(hyper)nuclei with a larger mass number \textit{A} are more sensitive to the baryochemical potential. The parameters of the SHM are extracted by performing a fit to the measured yields of different particle species~\cite{Andronic:2017pug} and using these extracted parameters, one can calculate the expected yields for other, yet unmeasured, particle species such as $A=4$ (anti)hypernuclei~\cite{Vovchenko:2019pjl}. 

An alternative approach to describe the production of nuclei is the coalescence model~\cite{Sun:2018mqq,ALICE:2015oer,Bellini:2018epz,Bellini:2019zqc}, which describes the formation of a nucleus by the coalescence of nucleons that are close in phase space at kinetic freeze-out. Modern coalescence calculations rely on the overlap between the  nuclear wave functions and the phase space distribution of the nucleons. In the coalescence model, the spin degeneracy is also taken into account~\cite{Scheibl:1998tk}. The production yield of the only $A=4$ (anti)nucleus observed so far, i.e. the ${}^{4}$He, is described within 0.3$\sigma$ by the SHM, while the current implementation of the coalescence model underestimates the data~\cite{ALICE:2023qyl}. There are no predictions yet for the production yields of (anti)hypernuclei with $A=4$ in any coalescence approach to compare to.

While the (anti)\hyp~$\Lambda$-separation energy is about 100 keV~\cite{massshift}, the larger $\Lambda$-separation energies of the $A=4$ (anti)hypernuclei presented in this Letter allow for the existence of excited states, which are known for many years. The most precise measurement of the excitation energy of the \hyperE~was carried out through gamma-ray spectroscopy at J-PARC~\cite{J-PARCE13:2015uwb} with an uncertainty of about 3.8\%. The excitation energy of the \hyper~was measured through stopped K$^{-}$ mesons in a ${}^{7}$Li target at CERN PS~\cite{Bedjidian:1976zh, 1979252}, for instance. The current world-average values for the ground-state binding energy ($J^{P}=0^{+}$) and energy of the excited state ($J^{P}=1^{+}$) are ($2.169 \pm 0.042$) MeV and ($1.081 \pm 0.046$) MeV for the \hyper, and ($2.347 \pm 0.036$) MeV and ($0.942 \pm 0.036$) MeV for the \hyperE, respectively~\cite{HypernuclearDataBase}. The known excited states decay electromagnetically to the ground state with lifetimes of the order of picoseconds~\cite{J-PARCE13:2015uwb,Bedjidian:1976zh}. Measurements of hypernuclei based on displaced decay vertex topologies originating from weak decays are only able to reconstruct the ground states. However, the decays of the excited states into the ground state contribute to the measured yield of the (anti)hypernuclei. In the framework of the SHM, the yields of the ground and excited states can be computed separately by taking into account their spin degeneracy through a factor of $2J+1$, where $J$ is the spin of the state~\cite{Donigus:2023hiw, Vovchenko_2020}. According to SHM calculations, the feed-down from the excited states is expected to enhance the yield of the two considered (anti)hypernuclei by a factor of four with respect to the case in which only the ground state is considered. The comparison of the measured yields with the SHM predictions could therefore provide information on the presence of such excited states.

As presented before, the ground-state binding energies and the masses of the (anti)\hyper ~and (anti)\hyperE ~differ by 178 keV~\cite{A1:2015isi,lh4br2}. The excitation energies also differ by 139 keV due to charge-symmetry breaking in (anti)hypernuclei~\cite{J-PARCE13:2015uwb,Bedjidian:1976zh}, which arises from the varying strengths of the hyperon--nucleon interactions between protons and neutrons~\cite{RevMod88:Strangeness,Haidenbauer:2021wld,Le:2022ikc}. Precision mass measurements of the (anti)\hyper, and (anti)\hyperE ~are needed to constrain the models for the description of hypernuclei, e.g. chiral effective field theories~\cite{Le:2023bfj,Haidenbauer:2023qhf}.  

The analysis of (anti)\hyper ~and (anti)\hyperE ~is performed using the 10\% most central (0--10\% centrality class) collisions of the Pb--Pb data sample collected by the ALICE Collaboration during the 2018 data-taking campaign at a center-of-mass energy per nucleon pair of $\sqrt{s_{\mathrm{NN}} }= 5.02$~TeV. The central Pb--Pb collisions used in the analysis are triggered online based on the signal of the V0 detector, which is composed of two scintillator arrays, called V0A and V0C, positioned at forward ($2.8  < \eta <  5.1$) and backward ($-3.7  < \eta <  -1.7$) pseudorapidities, respectively~\cite{ALICECollaboration_2008}. Minimum bias (MB) events are triggered by requiring coincident signals in V0A and V0C, while central collisions are selected by requiring large charge deposit in the V0 detector in addition to the MB trigger. Events are further selected offline by requiring that the position of the primary vertex along the beam axis is within $\pm$10 cm from the nominal interaction point, in order to ensure full geometrical acceptance in the Inner Tracking System (ITS) for $|\eta|<0.9$. After the event selections, a total of 108 million central Pb--Pb collisions are considered for the analysis.
For the reconstruction of the decay products of the (anti)hypernuclei, as well as the decay vertices, several subdetectors of the ALICE setup~\cite{ALICECollaboration_2008} are used. Track reconstruction in the central rapidity region is performed by using the ITS, the Time Projection Chamber (TPC), and the Time-Of-Flight Detector (TOF) which are embedded in a solenoid that provides a homogeneous magnetic field of 0.5 T. Charged-particle tracks are selected by requiring that they have $|\eta|<0.9$, at least 50 associated hits in the TPC, and a $\chi^{2}$/cluster for the momentum fit in the TPC lower than 4. Particle identification (PID) is carried out by utilizing the specific energy loss of particles inside the gas volume of the TPC, which is measured with a resolution of about 5\%~\cite{ALME2010316}. The specific energy loss as a function of the momentum can be described by a Bethe--Bloch parametrization~\cite{Blum:2008nqe} for each species. These parametrizations are used to select the (anti)hypernucleus decay products within a 4$\sigma$ interval, where the $\sigma$ is calculated through the resolution of the TPC. Additionally, the TOF detector allows for the measurement of the particle flight time, which provides an additional criterion for the identification of the charged particles. 

The (anti)hypernucleus candidates are constructed from a combination of two or three tracks that pass the track quality and PID selections as described above, and have the correct charge configuration. Selection criteria are applied to the reconstructed decay topologies to minimize the combinatorial background and enhance the signal purity. The same criteria are used for both hypernuclei and antihypernuclei. Raw yields are then extracted via an invariant-mass analysis of candidates passing the selection requirements. The fits to the invariant-mass spectra are conducted considering hypernuclei and antihypernuclei both together and separately. The reconstruction of the decay vertices utilizes a Kalman Filter approach~\cite{kfp2,kfp1,Gorbunov:2013yvt}, which allows to find decay vertices by combining tracks that likely originate from a common decay vertex. 

The selection of the (anti)hypernuclei exploits the displaced decay vertex topology. In particular, the distance of closest approach (DCA) of each track to the primary vertex, the DCA of each track to the decay vertex, the DCA between the tracks, the proper lifetime, and the cosine of the angle between the reconstructed momentum of the (anti)hypernucleus and the line connecting the primary and secondary vertices are determined for each (anti)hypernucleus candidate. In this analysis, the selections are based on a machine learning approach that takes into account the distributions of the previously mentioned track and vertex variables and the correlations among them. The approach used in this analysis utilizes a gradient boosted decision tree classifier (BDT)~\cite{xgboost, barioglio_2021_5734093}, which is trained using signal samples from the simulation and background samples from track combinations with the same charge sign (like-sign) taken from the data. For the two-body decay of the (anti)\hyper~ the like-sign sample is built from the like-sign (anti)${}^{4}$He and pion combinations, while for the three-body decay of the (anti)\hyperE ~the like-sign sample is built choosing the charge of the pion to be equal to one of the (anti)${}^{3}$He and the (anti)proton. The Monte Carlo (MC) sample is produced by using the HIJING event generator for the underlying Pb--Pb event~\cite{hijing}, while the (anti)hypernuclei are injected with uniform transverse momentum and rapidity distributions. The transport of the generated particles through the ALICE apparatus is done using GEANT3~\cite{geant3}, which simulates the interaction between the hypernucleus decay products and the detector material. The transverse momentum distribution of the simulated (anti)hypernuclei is reweighted afterwards using a Blast-Wave~\cite{blastwave2} distribution to obtain a more realistic description of the variables used for the selection. In heavy-ion collisions, the Blast-Wave model is a simplified parametrization of the hydrodynamical expansion of the medium, and provides a good description of the transverse momentum distributions of many particles including (anti)\hyp~\cite{ALICE:2015oer}. The parametrization of the Blast-Wave used in this analysis is taken from Ref.~\cite{ALICE:2023qyl} (Fit A). 

The machine learning procedure is split into two steps for this analysis. In the first step, the BDT is trained only on variables that are used for the particle identification of single tracks. In the second step, the training is performed on the variables that characterize the displaced decay vertex topology. This step-wise procedure helps to reject ${}^{3}$H misidentified as ${}^{3}$He in the (anti)\hyperE ~decay, since the particle identification using only the TPC specific energy loss information cannot separate ${}^{3}$H and ${}^{3}$He for momentum over charge $<1.0$  GeV/\textit{c}. The output of the BDT is a score which is proportional to the candidate probability of being signal. A threshold is then applied on the BDT output score to select the signal candidates. For the first machine learning step, this threshold is tuned to obtain a selection efficiency for the signal higher than 90\%. A selection is applied on the second machine learning step to maximise the statistical significance of the signal. To do so, a method based on the Punzi figure-of-merit (FoM)~\cite{punzi} is applied to maximize the inverse of the statistical uncertainty on the signal which corresponds to a maximization of the significance. This method provides a model-independent way to scan for the optimal result. 
The input values for this method are the geometrical detector acceptance and the reconstruction and selection efficiency (including tracking, vertexing, and BDT selection efficiencies). Additionally, the desired significance and the square-root of the number of background candidates in a 3$\sigma$ region around the hypernucleus mass is considered, where $\sigma$ is the Gaussian width of the signal invariant-mass peak. 

\begin{figure}[!h]
    \centering
    \includegraphics[width=0.48\linewidth]{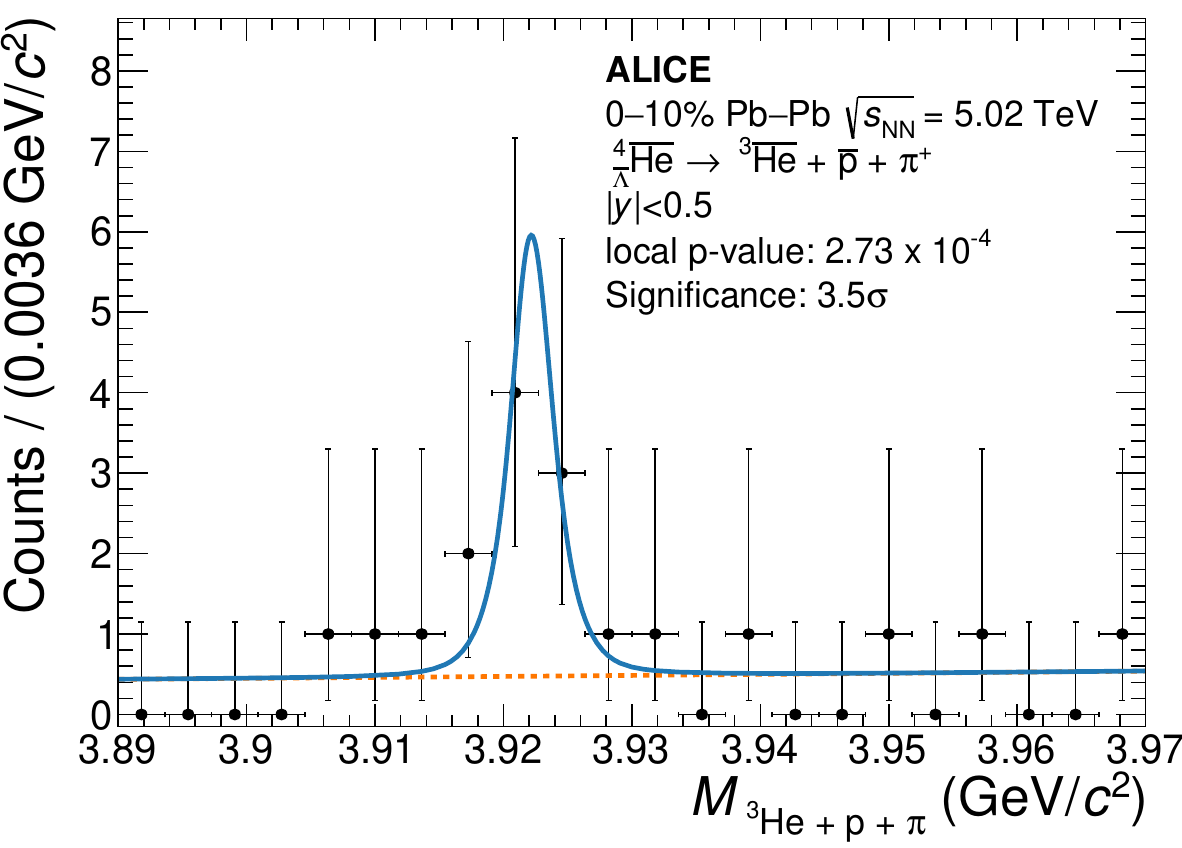}
    \includegraphics[width=0.48\linewidth]{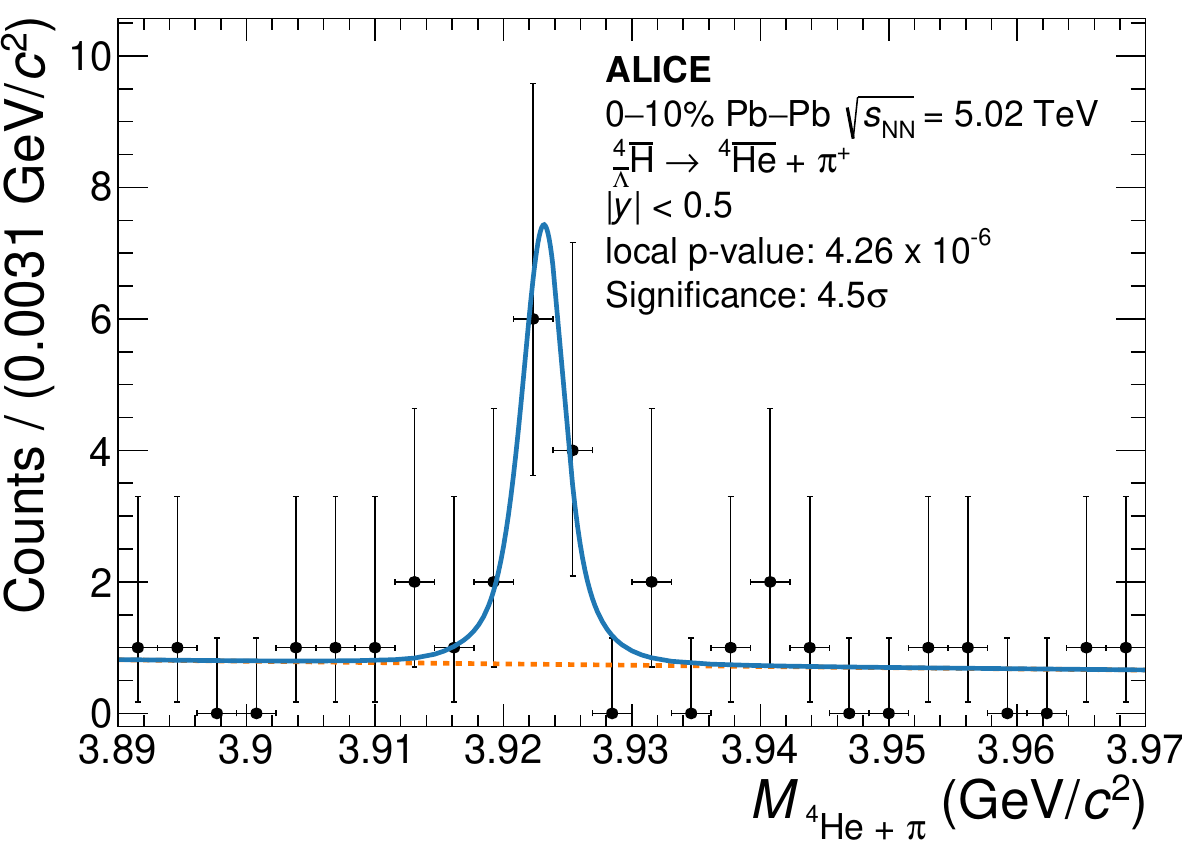}
    \caption{Invariant-mass spectra of \ahyperE ~candidates on the left and \ahyper ~candidates on the right. These spectra are fitted with a KDE template for the signal peak (blue curve) and an exponential function to model the background (orange curve). The combined invariant-mass spectra for the sum of particle and antiparticle states of each (anti)hypernucleus are shown in the End Matter.}
    \label{fig:invmassantiHe}
\end{figure}

After selection on the BDT output score, the signal is extracted by analyzing the invariant-mass spectra of the (anti)hypernuclei candidates. Figure~\ref{fig:invmassantiHe} shows the invariant-mass spectrum of the antihyperhelium-4 (\ahyperE) with a significance of 3.5$\sigma$ on the left, and the invariant-mass spectrum of the antihyperhydrogen-4 (\ahyper) with a significance of 4.5$\sigma$ on the right, while the invariant-mass spectra for the sum of the particle and antiparticle states of the (anti)\hyper ~and (anti)\hyperE~are presented in the End Matter. The invariant-mass spectrum of the \ahyperE~represents the first experimental evidence of this antihypernucleus. These spectra are fitted with a kernel density estimator function (KDE) for the signal~\cite{verkerke2003roofit, CRANMER2001198} (blue curve) and an exponential function to describe the background (orange curve). The KDE serves as a template, tuned to the MC signal shape, with Gaussian smoothing applied to regularize its shape. The corresponding significance is calculated with the asymptotic formulae for the profile likelihood ratio as described in Ref.~\cite{Cowan:2010js}, and expressed as a function of the invariant mass using the local p-value. 

In order to calculate the production yields per unit of rapidity, d\textit{N}/d\textit{y}, the raw counts are extracted by integrating the KDE signal functions obtained from the fits to the invariant-mass distributions of the sum of particle and antiparticle candidates reported in the End Matter. These counts are then normalized to the number of events in the analyzed data sample. They are divided by the width of the chosen rapidity region ($|y|<0.5$) and divided by two to obtain the production yields of particles from the raw counts, which are the sum of particles and antiparticles. The counts are then corrected for the geometrical acceptance and reconstruction and selection efficiency, the latter one being determined by applying the BDT model on the signal MC sample. Furthermore, the branching ratio of the specific decay mode is taken into account. For the (anti)\hyper, the branching ratio is assumed to be (55 $\pm$ 10)\%, which is based on a span of experimental estimates presented in Ref.~\cite{lh4br,lh4br2, Bedjidian:1976zh}. For the (anti)\hyperE, the branching ratio value of (28.9 $\pm$ 3.9)\% from the calculation of Parker \textit{et al.} reported in Ref.~\cite{Parker:2007yr}, is taken. Further corrections are applied for the absorption of the (anti)hypernuclei in the detector material, which are found to be 2.9\% and 3.5\% for the (anti)\hyper ~and the (anti)\hyperE, respectively. For this correction factor, a relative systematic uncertainty of 6.3\% for the (anti)\hyper ~and 5.5\% for the (anti)\hyperE ~is considered. 

Additionally, a systematic uncertainty to account for possible imperfections in the GEANT3 modeling of the absorption of the decay particles (about 4.5\% for each (anti)hypernucleus) is taken into account. For the systematic uncertainty on the BDT selection, different selections on the BDT output value are tested by varying the threshold by $\pm$ 5\% in steps of 1\% around the default value. The range of the variation of the BDT output selection was chosen to preserve a high significance for the signal and guarantee a robust extraction of the raw yield with the different selections. In addition, a systematic uncertainty on the input transverse-momentum shape of (anti)hypernuclei in the simulation is added, which is determined by varying the parameters of the Blast-Wave according to the values of the different fits from Ref.~\cite{ALICE:2023qyl}. Furthermore, the uncertainty on the yield extraction is considered, by varying the signal and background fit functions (using a Gaussian, Bukin~\cite{Bukin:2007zha}, or double-sided Crystal Ball function for the signal and different polynomial functions for the background description). The systematic uncertainties of the BDT selection, the raw yield extraction and the input transverse-momentum shape are assigned as the root-mean-square (rms) value of the distribution of the corrected yields from the aforementioned variations. The most dominant source of systematic uncertainty is the uncertainty on the branching ratio, which is about 18.2\% for the (anti)\hyper ~and about 13.5\% for the (anti)\hyperE. The total systematic uncertainty for the yield of the (anti)\hyper ~is about 22\% and for the (anti)\hyperE ~about 19\%. The values of all sources of systematic uncertainties and additional information on their evaluation can be found in the End Matter. 

In addition to the production yields, a measurement of the mass of the two investigated (anti)hypernuclei is performed. The mass is defined as the mean value of the signal fit function. However, the reconstructed value of the invariant mass is affected by the imperfect correction for the energy loss of the decay particles in the material of the ALICE detector. This effect produces a shift of the invariant-mass distribution that depends on the radial distance traveled by the (anti)hypernuclei before decaying, and the shift is corrected using the same procedure as Ref.~\cite{massshift}. The systematic uncertainty on the mass measurement is evaluated as the rms of the distribution of the mean values of the signal fit function obtained when varying the fit configuration as discussed above. Additional variations in the analysis procedure for the invariant-mass fit are considered in the estimation of the systematic uncertainty on the mass by varying the BDT selection and the input transverse-momentum shapes in the simulation. Both the contributions affect the templates of the signal invariant-mass peak extracted from the simulation.

\begin{figure}[!h]
    \centering
    \includegraphics[width=0.75\linewidth]{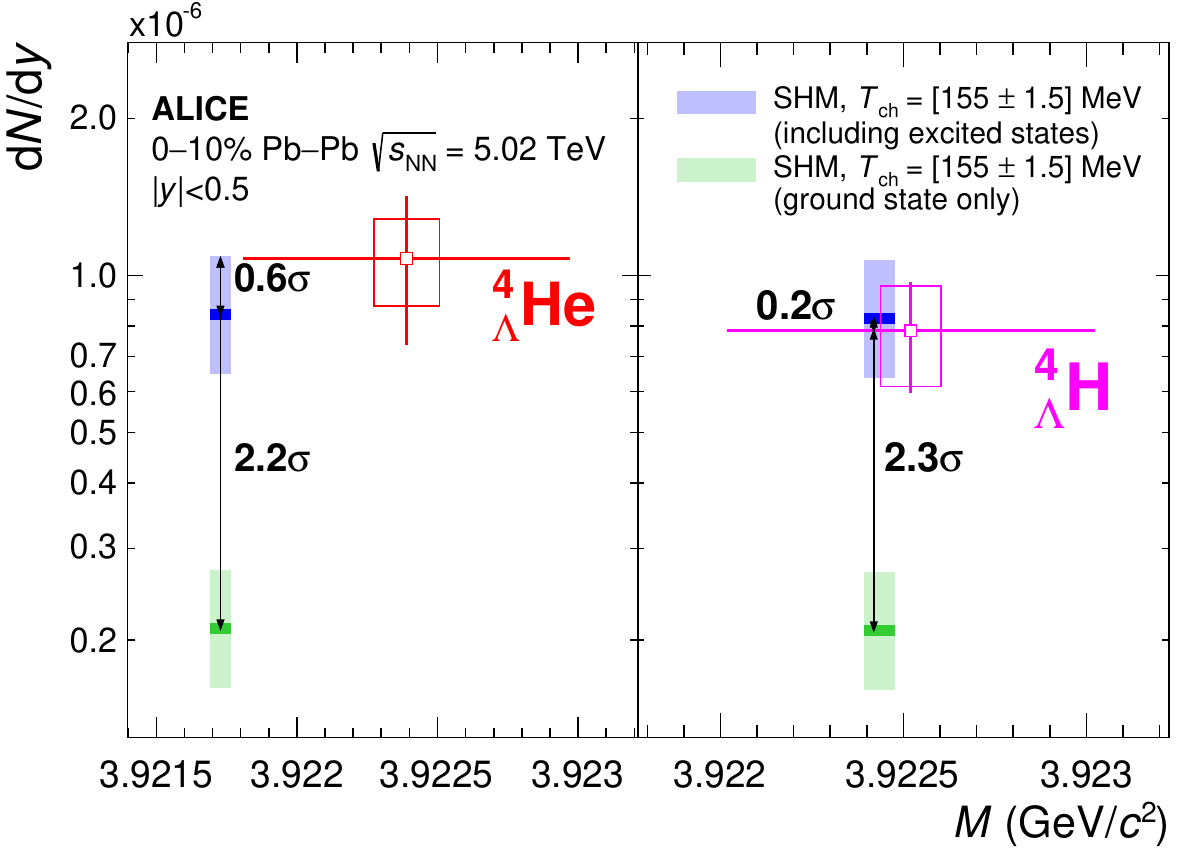}   
    \caption{Measured integrated yields of the (anti)\hyperE ~on the left and the (anti)\hyper ~on the right (average of particle and antiparticle state for each hypernucleus). The x-axis reports the measured mass, which is compared to the world-average values obtained from the Hypernuclei Database~\cite{HypernuclearDataBase}. The horizontal width of the blue and green lines corresponds to the uncertainty of the world-average values. The statistical uncertainties on the measured values are given by the bars and the systematic uncertainties by the boxes around the central values. A comparison is shown between the d\textit{N}/d\textit{y} values predicted by the SHM at $T_{\rm ch} = 155$ MeV, considering only the ground state (green line), and those including feed-down contributions from excited states (blue line) for each (anti)hypernucleus~\cite{Donigus:2023hiw, Vovchenko_2020, Vovchenko:2019pjl,Vovchenko:2018fiy}. The shaded areas around the predictions of the SHM correspond to the deviation of the expected yield by a variation of \textit{T}$_{\mathrm{ch}}$ of 1.5 MeV. The arrows connecting the experimental results and the d\textit{N}/d\textit{y} values predicted with the SHM (for $T_{\rm ch} = 155$ MeV) indicate their difference expressed as number of standard deviations using only the experimental uncertainties.}
    \label{fig:yields}
\end{figure}

The measured values of the yield per unit of rapidity, d\textit{N}/d\textit{y}, and the mass of the two considered (anti)hypernuclei with $A=4$ are shown in Fig.~\ref{fig:yields}. The results are compared to the predictions of the SHM at \textit{T}$_{\mathrm{ch}}=155$ MeV~\cite{Vovchenko:2019pjl,Vovchenko:2018fiy}. The green lines correspond to the SHM prediction with the ground state of the (anti)hypernuclei only, while the blue lines also include the feed-down contribution from the known excited states. Arrows in Fig.~\ref{fig:yields} illustrate the discrepancies between the measured d\textit{N}/d\textit{y} and the expectations. The shaded areas around the predictions of the SHM correspond to the deviation of the expected yield by a variation of \textit{T}$_{\mathrm{ch}}$ of 1.5 MeV. The difference between the ground-state-only SHM predictions and experimental data, expressed as number of standard deviations (calculated using the combination of statistical and systematical uncertainties), is found to be 2.3$\sigma$ and 2.2$\sigma$ for the (anti)\hyper ~and the (anti)\hyperE,~respectively. Combining these two significances using the Fisher's method for the combination of two independent p-values ~\cite{fisher,Fisher2}, a value of 3.3$\sigma$ is obtained. After including the excited states in the SHM predictions, the model agrees with the measured yields within 0.2$\sigma$ ~and 0.6$\sigma$ for the the (anti)\hyper ~and the (anti)\hyperE, respectively. 
Figure ~\ref{fig:yields} also shows the measurement of the mass for both (anti)hypernuclei, indicated by the horizontal position of the points, compared to the world-average values of previous mass measurements (blue and green lines), taken from the Hypernuclei Database~\cite{HypernuclearDataBase}. This database collects measurements of the properties of (anti)hypernuclei and calculates their world-average values, since these are not included into the PDG~\cite{Workman:2022ynf}. 

The presented measurement found the masses to be [3.9225 $\pm$ 0.0005 (stat.) $\pm$ 0.0001 (syst.)] GeV/\textit{c}$^{\mathrm{2}}$ for the (anti)\hyper ~and [3.9224 $\pm$ 0.0006 (stat.) $\pm$ 0.0001 (syst.)] GeV/\textit{c}$^{\mathrm{2}}$ for the (anti)\hyperE ~which are consistent with the world-average values within 1$\sigma$. The statistical uncertainties are dominant, while the systematic uncertainties are relatively small.

\newpage

\begin{figure}[!h]
    \centering
    \includegraphics[width=0.75\linewidth]{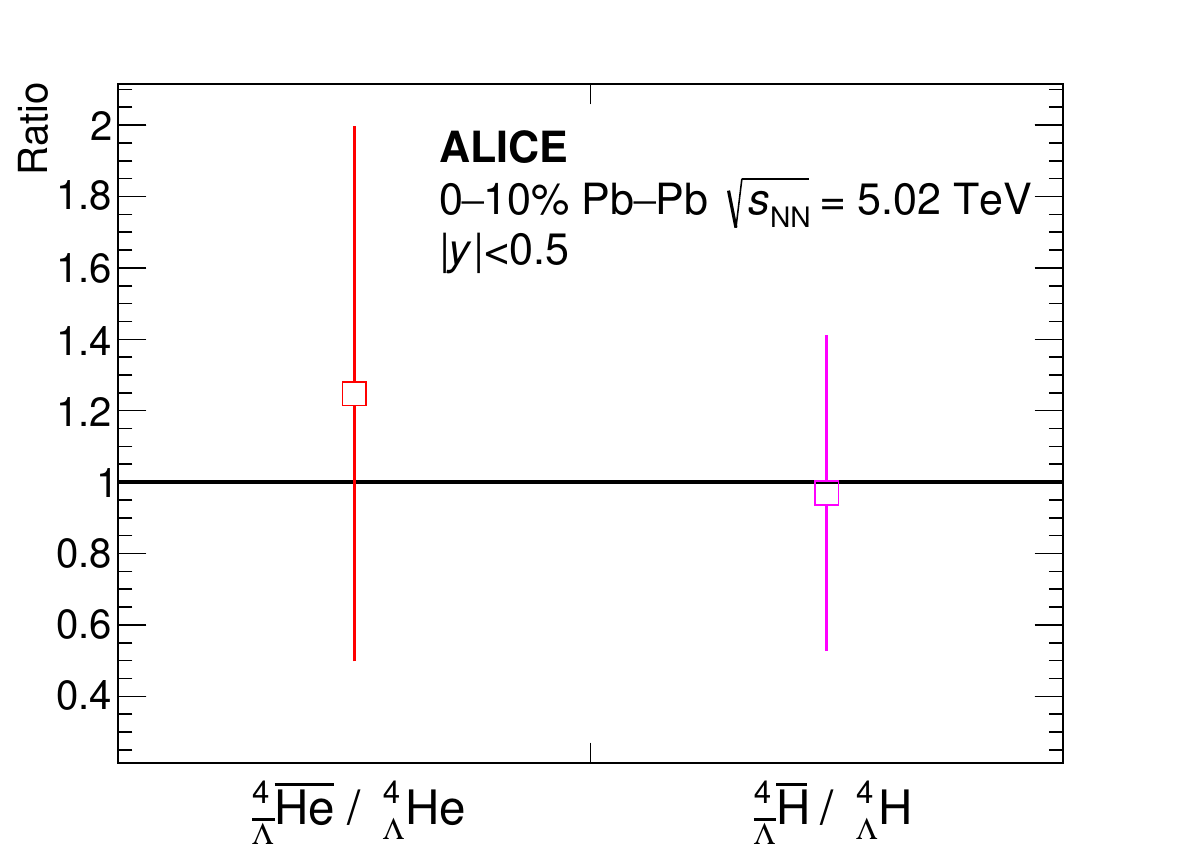}  
    \caption{Antiparticle-to-particle ratio for both investigated hypernuclei. The uncertainties are given by the quadratic sum of the statistical uncertainties and the systematic uncertainty on the absorption of the (anti)hypernuclei decay products (4.3\% for the \ahyperE / \hyperE, 4.5\% for the \ahyper / \hyper) and the absorption of the (anti)hypernuclei themselves (5.5\% for the \ahyperE / \hyperE, 6.3\% for the \ahyper / \hyper). }
    \label{fig:ratio}
\end{figure}

In addition, extracting the yields of the antihypernuclei separately, as reported in Fig.~\ref{fig:invmassantiHe}, allows for the evaluation of the antiparticle-to-particle ratios for both hypernuclei. The baryochemical potential, one of the three free parameters of the SHM, can be extracted from a fit to antiparticle-to-particle ratios~\cite{ALICE:2023ulv}. In fact, nuclei become more sensitive to this potential with increasing mass number $A$, making the two analyzed (anti)hypernuclei well-suited probes to test the predictions of the SHM. In Fig.~\ref{fig:ratio}, the antiparticle-to-particle ratios are shown for both hypernuclei. The obtained \ahyper / \hyper ~ratio is found to be $0.97 \; \pm \; 0.44$ and the \ahyperE / \hyperE ~ratio $1.25 \; \pm \; 0.75$, where the values of the uncertainties are the combination of the statistical and systematic uncertainties, since the systematic uncertainties are negligible. Since antiparticle and particle yields are obtained utilizing the same procedures, only the systematic uncertainty on the absorption of the (anti)hypernuclei decay products (4.3\% for the \ahyperE / \hyperE, 4.5\% for the \ahyper / \hyper) and the absorption of the (anti)hypernuclei themselves (5.5\% for the \ahyperE / \hyperE, 6.3\% for the \ahyper / \hyper) are taken into account for the ratio and added quadratically to the statistical uncertainties. The measurements agree with the SHM expectation where all antiparticle-to-particle ratios are close to one at midrapidity at LHC energies, at which the baryochemical potential is close to zero~\cite{ALICE:2023ulv}. 

\newpage

In this Letter, the first evidence of ${}^{4}_{\overline{\Lambda}}\overline{\textrm {He}}$ with a significance of $3.5\sigma$ is reported, along with the first measurement of the production of $A=4$ (anti)hypernuclei at the LHC. The measured (anti)\hyper~ and (anti)\hyperE ~mass values are compatible with the current world-average values. The measured production yields are compared with the predictions of the Statistical Hadronization Model at \textit{T}$_{\mathrm{ch}} = 155$ MeV. The comparison of the measured production yields with the calculations considering only the ground state reveals a deviation of $3.3\sigma$, while close agreement ($\leq 0.6\sigma$) is observed after including the feed-down from excited states. This confirms that the SHM is able to provide accurate predictions for the production yields of compact nuclear states such as (anti)hypernuclei in heavy-ion collisions and that the inclusion of the feed-down from excited states is crucial to describe the data. The presented measurements of the antiparticle-to-particle ratios are compatible with unity. This result is consistent with the latest measurement of the baryochemical potential published by ALICE, which is compatible with zero in heavy-ion collisions at LHC energies~\cite{ALICE:2023ulv}. 
The precision of the presented measurements of these rarely produced (anti)hypernuclei is limited by the available sample size, despite the systematic uncertainties being rather small. Previous measurements from the ALICE Collaboration on the (anti)\hyp~ mass and lifetime show that with a sufficient amount of candidates to analyze, very precise measurements on these weakly decaying particles are possible~\cite{massshift}. Precise measurements of the $A=4$ (anti)hypernuclei production and properties, enabled through larger data samples that will be collected with ALICE during the LHC Run 3 and Run 4, will help to constrain the models for the description of hypernuclei, e.g.~chiral effective field theories, in particular the charge-symmetry breaking effect~\cite{RevMod88:Strangeness,Le:2023bfj,Haidenbauer:2023qhf}.


\newenvironment{acknowledgement}{\relax}{\relax}
\begin{acknowledgement}
\section*{Acknowledgements}

The ALICE Collaboration would like to thank all its engineers and technicians for their invaluable contributions to the construction of the experiment and the CERN accelerator teams for the outstanding performance of the LHC complex.
The ALICE Collaboration gratefully acknowledges the resources and support provided by all Grid centres and the Worldwide LHC Computing Grid (WLCG) collaboration.
The ALICE Collaboration acknowledges the following funding agencies for their support in building and running the ALICE detector:
A. I. Alikhanyan National Science Laboratory (Yerevan Physics Institute) Foundation (ANSL), State Committee of Science and World Federation of Scientists (WFS), Armenia;
Austrian Academy of Sciences, Austrian Science Fund (FWF): [M 2467-N36] and Nationalstiftung f\"{u}r Forschung, Technologie und Entwicklung, Austria;
Ministry of Communications and High Technologies, National Nuclear Research Center, Azerbaijan;
Conselho Nacional de Desenvolvimento Cient\'{\i}fico e Tecnol\'{o}gico (CNPq), Financiadora de Estudos e Projetos (Finep), Funda\c{c}\~{a}o de Amparo \`{a} Pesquisa do Estado de S\~{a}o Paulo (FAPESP) and Universidade Federal do Rio Grande do Sul (UFRGS), Brazil;
Bulgarian Ministry of Education and Science, within the National Roadmap for Research Infrastructures 2020-2027 (object CERN), Bulgaria;
Ministry of Education of China (MOEC) , Ministry of Science \& Technology of China (MSTC) and National Natural Science Foundation of China (NSFC), China;
Ministry of Science and Education and Croatian Science Foundation, Croatia;
Centro de Aplicaciones Tecnol\'{o}gicas y Desarrollo Nuclear (CEADEN), Cubaenerg\'{\i}a, Cuba;
Ministry of Education, Youth and Sports of the Czech Republic, Czech Republic;
The Danish Council for Independent Research | Natural Sciences, the VILLUM FONDEN and Danish National Research Foundation (DNRF), Denmark;
Helsinki Institute of Physics (HIP), Finland;
Commissariat \`{a} l'Energie Atomique (CEA) and Institut National de Physique Nucl\'{e}aire et de Physique des Particules (IN2P3) and Centre National de la Recherche Scientifique (CNRS), France;
Bundesministerium f\"{u}r Bildung und Forschung (BMBF) and GSI Helmholtzzentrum f\"{u}r Schwerionenforschung GmbH, Germany;
General Secretariat for Research and Technology, Ministry of Education, Research and Religions, Greece;
National Research, Development and Innovation Office, Hungary;
Department of Atomic Energy Government of India (DAE), Department of Science and Technology, Government of India (DST), University Grants Commission, Government of India (UGC) and Council of Scientific and Industrial Research (CSIR), India;
National Research and Innovation Agency - BRIN, Indonesia;
Istituto Nazionale di Fisica Nucleare (INFN), Italy;
Japanese Ministry of Education, Culture, Sports, Science and Technology (MEXT) and Japan Society for the Promotion of Science (JSPS) KAKENHI, Japan;
Consejo Nacional de Ciencia (CONACYT) y Tecnolog\'{i}a, through Fondo de Cooperaci\'{o}n Internacional en Ciencia y Tecnolog\'{i}a (FONCICYT) and Direcci\'{o}n General de Asuntos del Personal Academico (DGAPA), Mexico;
Nederlandse Organisatie voor Wetenschappelijk Onderzoek (NWO), Netherlands;
The Research Council of Norway, Norway;
Pontificia Universidad Cat\'{o}lica del Per\'{u}, Peru;
Ministry of Science and Higher Education, National Science Centre and WUT ID-UB, Poland;
Korea Institute of Science and Technology Information and National Research Foundation of Korea (NRF), Republic of Korea;
Ministry of Education and Scientific Research, Institute of Atomic Physics, Ministry of Research and Innovation and Institute of Atomic Physics and Universitatea Nationala de Stiinta si Tehnologie Politehnica Bucuresti, Romania;
Ministry of Education, Science, Research and Sport of the Slovak Republic, Slovakia;
National Research Foundation of South Africa, South Africa;
Swedish Research Council (VR) and Knut \& Alice Wallenberg Foundation (KAW), Sweden;
European Organization for Nuclear Research, Switzerland;
Suranaree University of Technology (SUT), National Science and Technology Development Agency (NSTDA) and National Science, Research and Innovation Fund (NSRF via PMU-B B05F650021), Thailand;
Turkish Energy, Nuclear and Mineral Research Agency (TENMAK), Turkey;
National Academy of  Sciences of Ukraine, Ukraine;
Science and Technology Facilities Council (STFC), United Kingdom;
National Science Foundation of the United States of America (NSF) and United States Department of Energy, Office of Nuclear Physics (DOE NP), United States of America.
In addition, individual groups or members have received support from:
Czech Science Foundation (grant no. 23-07499S), Czech Republic;
FORTE project, reg.\ no.\ CZ.02.01.01/00/22\_008/0004632, Czech Republic, co-funded by the European Union, Czech Republic;
European Research Council (grant no. 950692), European Union;
ICSC - Centro Nazionale di Ricerca in High Performance Computing, Big Data and Quantum Computing, European Union - NextGenerationEU;
Academy of Finland (Center of Excellence in Quark Matter) (grant nos. 346327, 346328), Finland.

\end{acknowledgement}

\bibliographystyle{utphys}   
\bibliography{bibliography}

\newpage
\appendix

\section{End Matter}

\subsection*{Invariant-mass spectra}
\begin{figure}[!h]
    \centering    
    \includegraphics[width=0.45\linewidth]{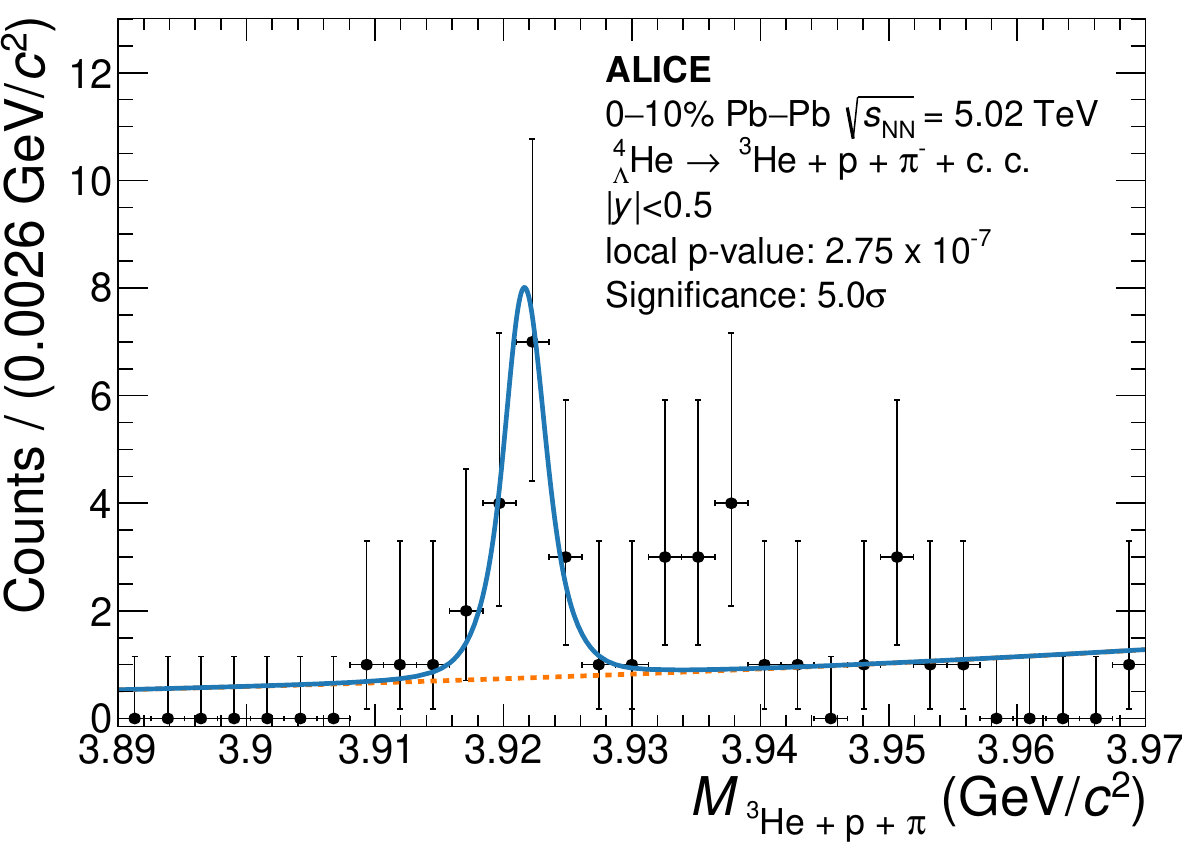}
    \includegraphics[width=0.45\linewidth]{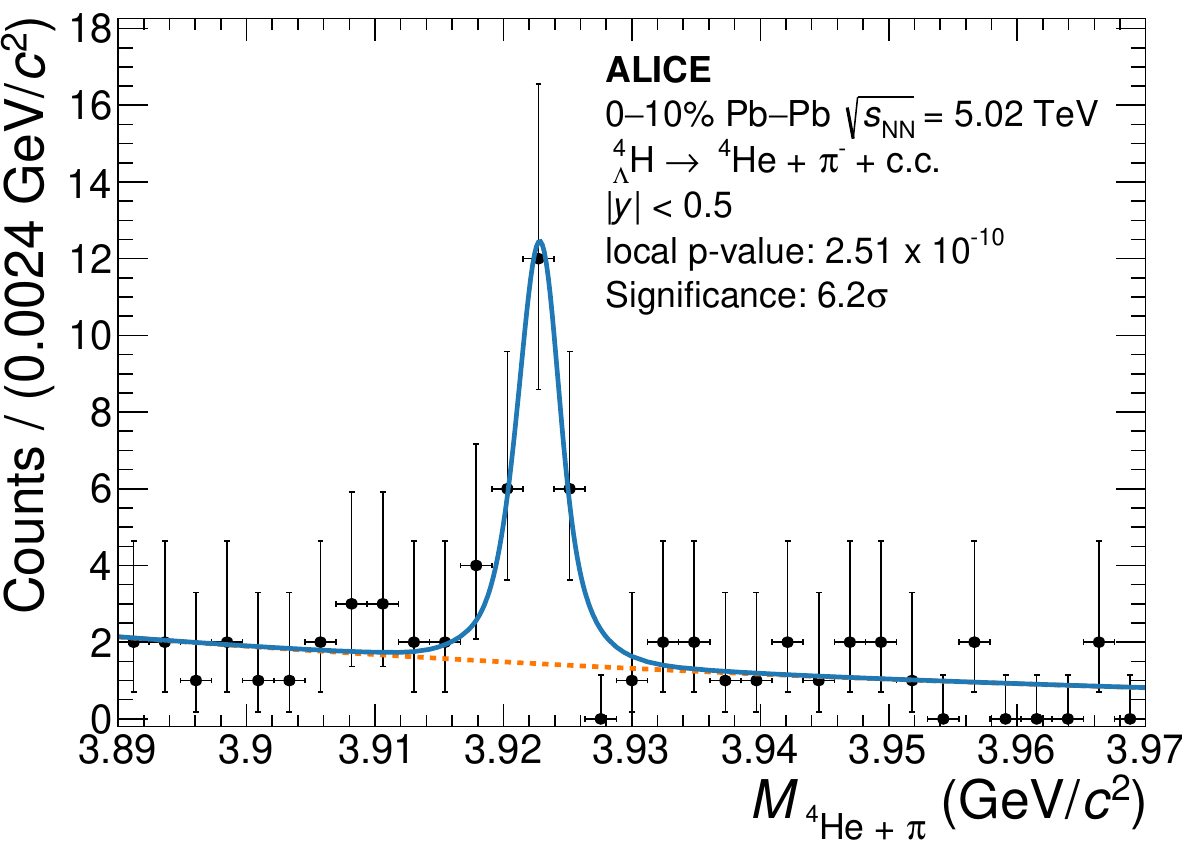}
    \includegraphics[width=0.45\linewidth]{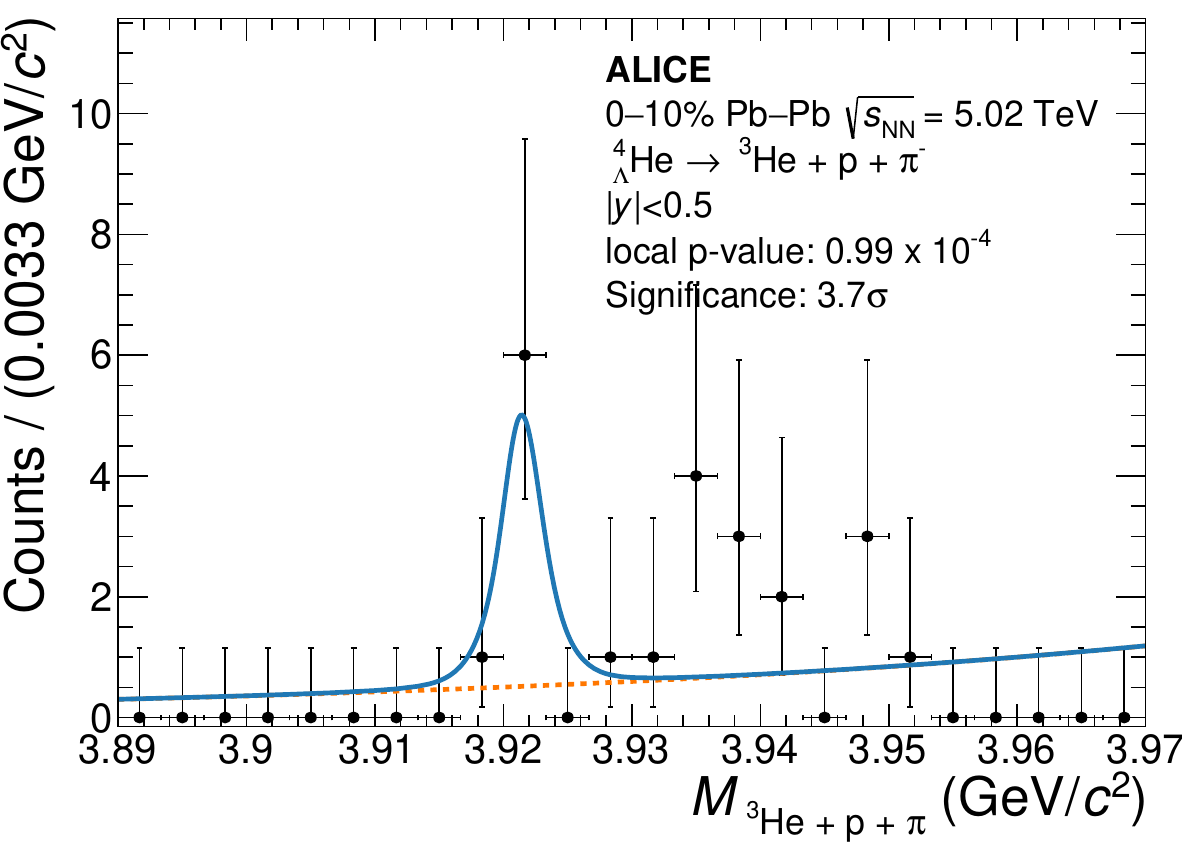}
    \includegraphics[width=0.45\linewidth]{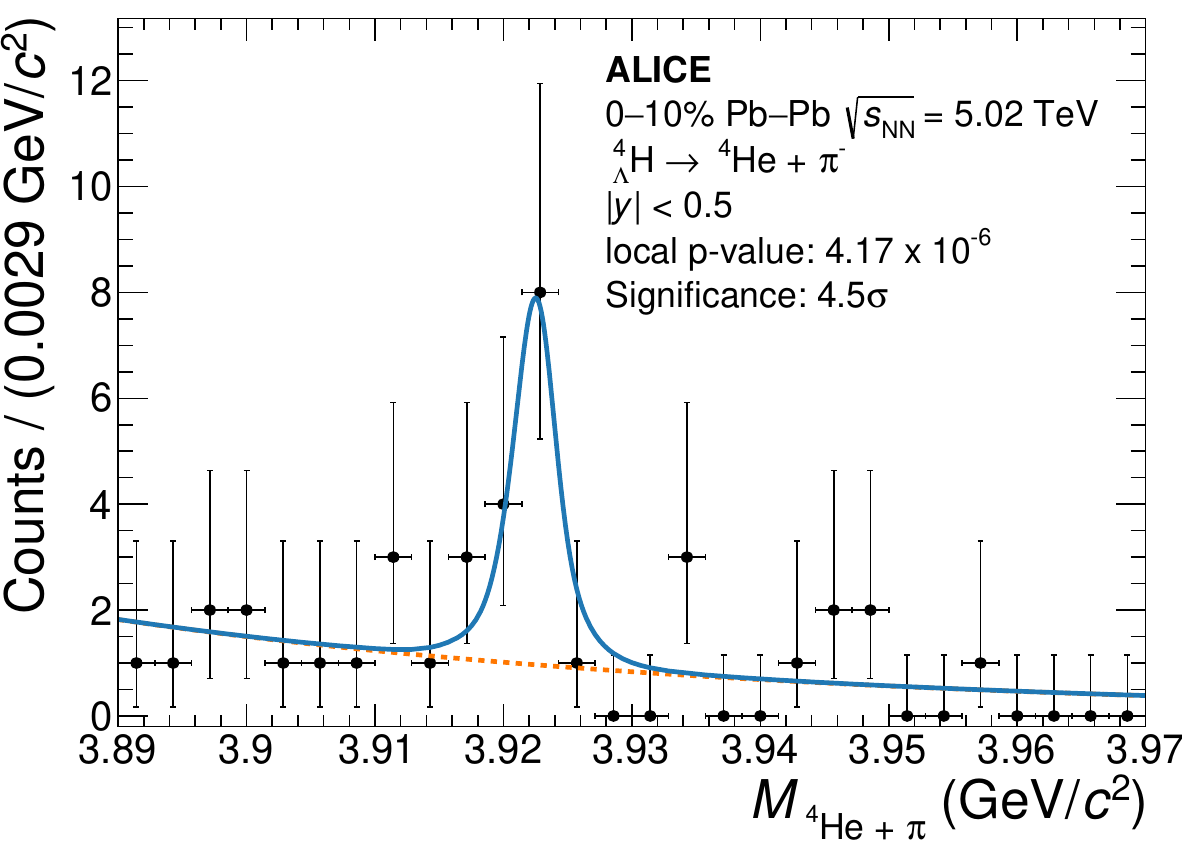}
    \caption{Invariant-mass spectra of the sum of particle and antiparticle states of (anti)\hyperE ~candidates on the upper left and of (anti)\hyper ~candidates on the upper right. In the lower panels, the invariant-mass spectra of the particle fraction only of the \hyperE ~candidates on the left and of the \hyper ~candidates on the right are presented. Each spectrum is fitted with a KDE template for the signal peak (blue curve) and an exponential function to model the background (orange curve).}
    \label{fig:invmass}
\end{figure}

The top panels of Fig. \ref{fig:invmass} show the invariant-mass spectra of the sum of particle and antiparticle states of the (anti)\hyper ~and (anti)\hyperE ~candidates, while the lower panels present the invariant-mass spectra of the particle fraction only. These spectra are selected and fitted as described in the text of the Letter. 

\subsection*{Details on the systematic uncertainty evaluation}
As described in the Letter, systematic uncertainties on the branching ratio, the description of the signal and background of the invariant-spectra, the BDT selection, the absorption of the decay products and the (anti)hypernuclei themselves as well as the input transverse-momentum shape are considered for the measurement of the corrected yield per unit of rapidity of the investigated (anti)hypernuclei. In the following, additional information on the determination of some of these systematic uncertainties is given and Table \ref{final_valuncert_hypnuc_new_IRC} provides an overview of the systematic uncertainties estimated for the d\textit{N}/d\textit{y} measurements (values are rounded).
\smallskip
\\The systematic uncertainty due to possible imperfections in the description of the absorption of the (anti)hypernucleus decay products in the simulation was estimated as the difference between the GEANT3 (modified in the ALICE software framework including absorption estimates~\cite{ALICE:2017xrp, ALICE:2015wav}) results and those obtained from a smaller sample simulated with the GEANT4 transport code~\cite{GEANT4:2002zbu}. Both models use different approaches to calculate the absorption of (anti)alpha and (anti)helium-3. However, the GEANT4 has been confirmed to describe the data for antihelium-3, while the antialpha absorption is not well known~\cite{geant4, ALICE:2023qyl}. The estimation of the systematic uncertainty results in a value of about 4.5\%. 
\smallskip
\\The fraction of $A=4$ (anti)hypernuclei which are absorbed in the ALICE detector material can not be extracted directly from the available simulation, as the (anti)\hyper ~and the (anti)\hyperE ~inelastic cross sections have never been measured. As an estimate, the absorption cross section of the ${}^{4}\overline{\mathrm{He}}$~\cite{ALICE:2023qyl} is used and scaled by the relative difference of the radii of each (anti)hypernucleus and of the ${}^{4}\overline{\mathrm{He}}$, respectively~\cite{geant4}. However, the absorption correction depends on the amount of material crossed by the (anti)hypernuclei and it is therefore different for (anti)\hyper ~and the (anti)\hyperE, which have different average decay lengths (the lifetimes are 250 picoseconds and 208 picoseconds for the (anti)\hyperE~and (anti)\hyper, respectively~\cite{HypernuclearDataBase}).  In particular, the correction factors are 2.9\% and 3.5\% for the (anti)\hyper ~and (anti)\hyperE, respectively. The systematic uncertainty on these correction factors is evaluated from the uncertainty of the (anti)(hyper)nuclei radii that enter the correction estimation. This leads to a relative uncertainty of 6.3\% for the (anti)\hyper ~and 5.5\% for the (anti)\hyperE. 
\smallskip
\\Another systematic uncertainty is added for the input transverse-momentum shape of \mbox{(anti)hypernuclei} in the simulation. This input transverse-momentum shape affects the distributions of the variables used in the BDT selections and the correction factors for the geometrical acceptance and reconstruction efficiency. For this purpose, the four different parameter sets for the Blast-Wave as described in Ref.~\cite{ALICE:2023qyl} are used as alternative input shapes and the resulting differences (rms) in the corrected yields are considered for the estimation of the systematic uncertainty. 

\begin{table}[!tbh]
\centering
\caption{Overview of systematic uncertainties for the d$N$/d$y$ of (anti)\hyper ~and (anti)\hyperE.}
\begin{tabular}{|c|c|c|}    
\hline
\textbf{Systematic Uncertainty} & \textbf{(anti)${}^{4}_{\Lambda}$H} & \textbf{(anti)${}^{4}_{\Lambda}$He} \\
\hline
 Branching Ratio & 18.2\% & 13.5\% \\
\hline
Raw yield extraction & 3.4\% & 4.3\% \\
\hline
BDT selection & 4.7\% & 1.7\% \\
\hline
Decay product absorption & 4.5\% & 4.3\% \\
\hline
(Anti)hypernucleus absorption & 6.3\% & 5.5\% \\
\hline
Input \textit{p}$_{\mathrm T}$-shape & 7.3\% & 10.3\% \\
\hline
Total & 22\% & 19\% \\
\hline
\end{tabular}
\label{final_valuncert_hypnuc_new_IRC}
\end{table}

\clearpage

%
%

\section{The ALICE Collaboration}
\label{app:collab}
\begin{flushleft} 
\small

S.~Acharya\,\orcidlink{0000-0002-9213-5329}\,$^{\rm 127}$, 
A.~Agarwal$^{\rm 135}$, 
G.~Aglieri Rinella\,\orcidlink{0000-0002-9611-3696}\,$^{\rm 32}$, 
L.~Aglietta\,\orcidlink{0009-0003-0763-6802}\,$^{\rm 24}$, 
M.~Agnello\,\orcidlink{0000-0002-0760-5075}\,$^{\rm 29}$, 
N.~Agrawal\,\orcidlink{0000-0003-0348-9836}\,$^{\rm 25}$, 
Z.~Ahammed\,\orcidlink{0000-0001-5241-7412}\,$^{\rm 135}$, 
S.~Ahmad\,\orcidlink{0000-0003-0497-5705}\,$^{\rm 15}$, 
S.U.~Ahn\,\orcidlink{0000-0001-8847-489X}\,$^{\rm 71}$, 
I.~Ahuja\,\orcidlink{0000-0002-4417-1392}\,$^{\rm 36}$
% ...

\section*{Affiliation Notes}

$^{\rm I}$ Deceased\\
$^{\rm II}$ Also at: Max-Planck-Institut fur Physik, Munich, Germany\\
$^{\rm III}$ Also at: Italian National Agency for New Technologies, Energy and Sustainable Economic Development (ENEA), Bologna, Italy\\
$^{\rm IV}$ Also at: Dipartimento DET del Politecnico di Torino, Turin, Italy\\
$^{\rm V}$ Also at: Department of Applied Physics, Aligarh Muslim University, Aligarh, India\\
$^{\rm VI}$ Also at: Institute of Theoretical Physics, University of Wroclaw, Poland\\
$^{\rm VII}$ Also at: Facultad de Ciencias, Universidad Nacional Autónoma de México, Mexico City, Mexico\\
$^{\rm VIII}$ Also at: An institution covered by a cooperation agreement with CERN\\

\section*{Collaboration Institutes}

$^{1}$ A.I. Alikhanyan National Science Laboratory (Yerevan Physics Institute) Foundation, Yerevan, Armenia\\
$^{2}$ AGH University of Krakow, Cracow, Poland\\
$^{3}$ Bogolyubov Institute for Theoretical Physics, National Academy of Sciences of Ukraine, Kiev, Ukraine\\
$^{4}$ Bose Institute, Department of Physics  and Centre for Astroparticle Physics and Space Science (CAPSS), Kolkata, India\\
$^{5}$ California Polytechnic State University, San Luis Obispo, California, United States\\
$^{6}$ Central China Normal University, Wuhan, China\\
$^{7}$ Centro de Aplicaciones Tecnol\'{o}gicas y Desarrollo Nuclear (CEADEN), Havana, Cuba\\
$^{8}$ Centro de Investigaci\'{o}n y de Estudios Avanzados (CINVESTAV), Mexico City and M\'{e}rida, Mexico\\
$^{9}$ Chicago State University, Chicago, Illinois, United States\\
$^{10}$ China Institute of Atomic Energy, Beijing, China\\
$^{11}$ China University of Geosciences, Wuhan, China\\
$^{12}$ Chungbuk National University, Cheongju, Republic of Korea\\
$^{13}$ Comenius University Bratislava, Faculty of Mathematics, Physics and Informatics, Bratislava, Slovak Republic\\
$^{14}$ Creighton University, Omaha, Nebraska, United States\\
$^{15}$ Department of Physics, Aligarh Muslim University, Aligarh, India\\
$^{16}$ Department of Physics, Pusan National University, Pusan, Republic of Korea\\
$^{17}$ Department of Physics, Sejong University, Seoul, Republic of Korea\\
$^{18}$ Department of Physics, University of California, Berkeley, California, United States\\
$^{19}$ Department of Physics, University of Oslo, Oslo, Norway\\
$^{20}$ Department of Physics and Technology, University of Bergen, Bergen, Norway\\
$^{21}$ Dipartimento di Fisica, Universit\`{a} di Pavia, Pavia, Italy\\
$^{22}$ Dipartimento di Fisica dell'Universit\`{a} and Sezione INFN, Cagliari, Italy\\
$^{23}$ Dipartimento di Fisica dell'Universit\`{a} and Sezione INFN, Trieste, Italy\\
$^{24}$ Dipartimento di Fisica dell'Universit\`{a} and Sezione INFN, Turin, Italy\\
$^{25}$ Dipartimento di Fisica e Astronomia dell'Universit\`{a} and Sezione INFN, Bologna, Italy\\
$^{26}$ Dipartimento di Fisica e Astronomia dell'Universit\`{a} and Sezione INFN, Catania, Italy\\
$^{27}$ Dipartimento di Fisica e Astronomia dell'Universit\`{a} and Sezione INFN, Padova, Italy\\
$^{28}$ Dipartimento di Fisica `E.R.~Caianiello' dell'Universit\`{a} and Gruppo Collegato INFN, Salerno, Italy\\
$^{29}$ Dipartimento DISAT del Politecnico and Sezione INFN, Turin, Italy\\
$^{30}$ Dipartimento di Scienze MIFT, Universit\`{a} di Messina, Messina, Italy\\
$^{31}$ Dipartimento Interateneo di Fisica `M.~Merlin' and Sezione INFN, Bari, Italy\\
$^{32}$ European Organization for Nuclear Research (CERN), Geneva, Switzerland\\
$^{33}$ Faculty of Electrical Engineering, Mechanical Engineering and Naval Architecture, University of Split, Split, Croatia\\
$^{34}$ Faculty of Nuclear Sciences and Physical Engineering, Czech Technical University in Prague, Prague, Czech Republic\\
$^{35}$ Faculty of Physics, Sofia University, Sofia, Bulgaria\\
$^{36}$ Faculty of Science, P.J.~\v{S}af\'{a}rik University, Ko\v{s}ice, Slovak Republic\\
$^{37}$ Faculty of Technology, Environmental and Social Sciences, Bergen, Norway\\
$^{38}$ Frankfurt Institute for Advanced Studies, Johann Wolfgang Goethe-Universit\"{a}t Frankfurt, Frankfurt, Germany\\
$^{39}$ Fudan University, Shanghai, China\\
$^{40}$ Gangneung-Wonju National University, Gangneung, Republic of Korea\\
$^{41}$ Gauhati University, Department of Physics, Guwahati, India\\
$^{42}$ Helmholtz-Institut f\"{u}r Strahlen- und Kernphysik, Rheinische Friedrich-Wilhelms-Universit\"{a}t Bonn, Bonn, Germany\\
$^{43}$ Helsinki Institute of Physics (HIP), Helsinki, Finland\\
$^{44}$ High Energy Physics Group,  Universidad Aut\'{o}noma de Puebla, Puebla, Mexico\\
$^{45}$ Horia Hulubei National Institute of Physics and Nuclear Engineering, Bucharest, Romania\\
$^{46}$ HUN-REN Wigner Research Centre for Physics, Budapest, Hungary\\
$^{47}$ Indian Institute of Technology Bombay (IIT), Mumbai, India\\
$^{48}$ Indian Institute of Technology Indore, Indore, India\\
$^{49}$ INFN, Laboratori Nazionali di Frascati, Frascati, Italy\\
$^{50}$ INFN, Sezione di Bari, Bari, Italy\\
$^{51}$ INFN, Sezione di Bologna, Bologna, Italy\\
$^{52}$ INFN, Sezione di Cagliari, Cagliari, Italy\\
$^{53}$ INFN, Sezione di Catania, Catania, Italy\\
$^{54}$ INFN, Sezione di Padova, Padova, Italy\\
$^{55}$ INFN, Sezione di Pavia, Pavia, Italy\\
$^{56}$ INFN, Sezione di Torino, Turin, Italy\\
$^{57}$ INFN, Sezione di Trieste, Trieste, Italy\\
$^{58}$ Inha University, Incheon, Republic of Korea\\
$^{59}$ Institute for Gravitational and Subatomic Physics (GRASP), Utrecht University/Nikhef, Utrecht, Netherlands\\
$^{60}$ Institute of Experimental Physics, Slovak Academy of Sciences, Ko\v{s}ice, Slovak Republic\\
$^{61}$ Institute of Physics, Homi Bhabha National Institute, Bhubaneswar, India\\
$^{62}$ Institute of Physics of the Czech Academy of Sciences, Prague, Czech Republic\\
$^{63}$ Institute of Space Science (ISS), Bucharest, Romania\\
$^{64}$ Institut f\"{u}r Kernphysik, Johann Wolfgang Goethe-Universit\"{a}t Frankfurt, Frankfurt, Germany\\
$^{65}$ Instituto de Ciencias Nucleares, Universidad Nacional Aut\'{o}noma de M\'{e}xico, Mexico City, Mexico\\
$^{66}$ Instituto de F\'{i}sica, Universidade Federal do Rio Grande do Sul (UFRGS), Porto Alegre, Brazil\\
$^{67}$ Instituto de F\'{\i}sica, Universidad Nacional Aut\'{o}noma de M\'{e}xico, Mexico City, Mexico\\
$^{68}$ iThemba LABS, National Research Foundation, Somerset West, South Africa\\
$^{69}$ Jeonbuk National University, Jeonju, Republic of Korea\\
$^{70}$ Johann-Wolfgang-Goethe Universit\"{a}t Frankfurt Institut f\"{u}r Informatik, Fachbereich Informatik und Mathematik, Frankfurt, Germany\\
$^{71}$ Korea Institute of Science and Technology Information, Daejeon, Republic of Korea\\
$^{72}$ KTO Karatay University, Konya, Turkey\\
$^{73}$ Laboratoire de Physique Subatomique et de Cosmologie, Universit\'{e} Grenoble-Alpes, CNRS-IN2P3, Grenoble, France\\
$^{74}$ Lawrence Berkeley National Laboratory, Berkeley, California, United States\\
$^{75}$ Lund University Department of Physics, Division of Particle Physics, Lund, Sweden\\
$^{76}$ Nagasaki Institute of Applied Science, Nagasaki, Japan\\
$^{77}$ Nara Women{'}s University (NWU), Nara, Japan\\
$^{78}$ National and Kapodistrian University of Athens, School of Science, Department of Physics , Athens, Greece\\
$^{79}$ National Centre for Nuclear Research, Warsaw, Poland\\
$^{80}$ National Institute of Science Education and Research, Homi Bhabha National Institute, Jatni, India\\
$^{81}$ National Nuclear Research Center, Baku, Azerbaijan\\
$^{82}$ National Research and Innovation Agency - BRIN, Jakarta, Indonesia\\
$^{83}$ Niels Bohr Institute, University of Copenhagen, Copenhagen, Denmark\\
$^{84}$ Nikhef, National institute for subatomic physics, Amsterdam, Netherlands\\
$^{85}$ Nuclear Physics Group, STFC Daresbury Laboratory, Daresbury, United Kingdom\\
$^{86}$ Nuclear Physics Institute of the Czech Academy of Sciences, Husinec-\v{R}e\v{z}, Czech Republic\\
$^{87}$ Oak Ridge National Laboratory, Oak Ridge, Tennessee, United States\\
$^{88}$ Ohio State University, Columbus, Ohio, United States\\
$^{89}$ Physics department, Faculty of science, University of Zagreb, Zagreb, Croatia\\
$^{90}$ Physics Department, Panjab University, Chandigarh, India\\
$^{91}$ Physics Department, University of Jammu, Jammu, India\\
$^{92}$ Physics Program and International Institute for Sustainability with Knotted Chiral Meta Matter (WPI-SKCM$^{2}$), Hiroshima University, Hiroshima, Japan\\
$^{93}$ Physikalisches Institut, Eberhard-Karls-Universit\"{a}t T\"{u}bingen, T\"{u}bingen, Germany\\
$^{94}$ Physikalisches Institut, Ruprecht-Karls-Universit\"{a}t Heidelberg, Heidelberg, Germany\\
$^{95}$ Physik Department, Technische Universit\"{a}t M\"{u}nchen, Munich, Germany\\
$^{96}$ Politecnico di Bari and Sezione INFN, Bari, Italy\\
$^{97}$ Research Division and ExtreMe Matter Institute EMMI, GSI Helmholtzzentrum f\"ur Schwerionenforschung GmbH, Darmstadt, Germany\\
$^{98}$ Saga University, Saga, Japan\\
$^{99}$ Saha Institute of Nuclear Physics, Homi Bhabha National Institute, Kolkata, India\\
$^{100}$ School of Physics and Astronomy, University of Birmingham, Birmingham, United Kingdom\\
$^{101}$ Secci\'{o}n F\'{\i}sica, Departamento de Ciencias, Pontificia Universidad Cat\'{o}lica del Per\'{u}, Lima, Peru\\
$^{102}$ Stefan Meyer Institut f\"{u}r Subatomare Physik (SMI), Vienna, Austria\\
$^{103}$ SUBATECH, IMT Atlantique, Nantes Universit\'{e}, CNRS-IN2P3, Nantes, France\\
$^{104}$ Sungkyunkwan University, Suwon City, Republic of Korea\\
$^{105}$ Suranaree University of Technology, Nakhon Ratchasima, Thailand\\
$^{106}$ Technical University of Ko\v{s}ice, Ko\v{s}ice, Slovak Republic\\
$^{107}$ The Henryk Niewodniczanski Institute of Nuclear Physics, Polish Academy of Sciences, Cracow, Poland\\
$^{108}$ The University of Texas at Austin, Austin, Texas, United States\\
$^{109}$ Universidad Aut\'{o}noma de Sinaloa, Culiac\'{a}n, Mexico\\
$^{110}$ Universidade de S\~{a}o Paulo (USP), S\~{a}o Paulo, Brazil\\
$^{111}$ Universidade Estadual de Campinas (UNICAMP), Campinas, Brazil\\
$^{112}$ Universidade Federal do ABC, Santo Andre, Brazil\\
$^{113}$ Universitatea Nationala de Stiinta si Tehnologie Politehnica Bucuresti, Bucharest, Romania\\
$^{114}$ University of Cape Town, Cape Town, South Africa\\
$^{115}$ University of Derby, Derby, United Kingdom\\
$^{116}$ University of Houston, Houston, Texas, United States\\
$^{117}$ University of Jyv\"{a}skyl\"{a}, Jyv\"{a}skyl\"{a}, Finland\\
$^{118}$ University of Kansas, Lawrence, Kansas, United States\\
$^{119}$ University of Liverpool, Liverpool, United Kingdom\\
$^{120}$ University of Science and Technology of China, Hefei, China\\
$^{121}$ University of South-Eastern Norway, Kongsberg, Norway\\
$^{122}$ University of Tennessee, Knoxville, Tennessee, United States\\
$^{123}$ University of the Witwatersrand, Johannesburg, South Africa\\
$^{124}$ University of Tokyo, Tokyo, Japan\\
$^{125}$ University of Tsukuba, Tsukuba, Japan\\
$^{126}$ Universit\"{a}t M\"{u}nster, Institut f\"{u}r Kernphysik, M\"{u}nster, Germany\\
$^{127}$ Universit\'{e} Clermont Auvergne, CNRS/IN2P3, LPC, Clermont-Ferrand, France\\
$^{128}$ Universit\'{e} de Lyon, CNRS/IN2P3, Institut de Physique des 2 Infinis de Lyon, Lyon, France\\
$^{129}$ Universit\'{e} de Strasbourg, CNRS, IPHC UMR 7178, F-67000 Strasbourg, France, Strasbourg, France\\
$^{130}$ Universit\'{e} Paris-Saclay, Centre d'Etudes de Saclay (CEA), IRFU, D\'{e}partment de Physique Nucl\'{e}aire (DPhN), Saclay, France\\
$^{131}$ Universit\'{e}  Paris-Saclay, CNRS/IN2P3, IJCLab, Orsay, France\\
$^{132}$ Universit\`{a} degli Studi di Foggia, Foggia, Italy\\
$^{133}$ Universit\`{a} del Piemonte Orientale, Vercelli, Italy\\
$^{134}$ Universit\`{a} di Brescia, Brescia, Italy\\
$^{135}$ Variable Energy Cyclotron Centre, Homi Bhabha National Institute, Kolkata, India\\
$^{136}$ Warsaw University of Technology, Warsaw, Poland\\
$^{137}$ Wayne State University, Detroit, Michigan, United States\\
$^{138}$ Yale University, New Haven, Connecticut, United States\\
$^{139}$ Yonsei University, Seoul, Republic of Korea\\
$^{140}$ Affiliated with an institute covered by a cooperation agreement with CERN\\
$^{141}$ Affiliated with an international laboratory covered by a cooperation agreement with CERN.\\

\end{flushleft} 

\end{document}